\documentclass[preprint,3p,times]{elsarticle}
\usepackage[british]{babel}
\usepackage{amsmath,amssymb}
\usepackage{textcomp}
\usepackage{graphicx}
\usepackage{booktabs}
\usepackage{array}
\usepackage{lineno}
\usepackage{url}
\usepackage[hidelinks]{hyperref}

\newcommand{\aifigdata}{Drawn with AI-assisted code from the authors' data.}
\newcommand{\aifigdesign}{Drawn with AI-assisted code from the authors' design.}
\newsavebox{\figbox}
\newcommand{\fitfigure}[1]{\sbox{\figbox}{\includegraphics{#1}}\ifdim\wd\figbox>\textwidth\includegraphics[width=\textwidth]{#1}\else\usebox{\figbox}\fi}
\newcommand{\us}{\textmu s}
\biboptions{sort&compress}
\journal{Measurement}

\begin{document}

\begin{frontmatter}

\title{Wideband overlapped polyphase channelizer for kinetic inductance detector readout: split-path architecture and experimental characterization}

\author[a,b]{Alberto Hern\'andez Fern\'andez\corref{cor1}}
\ead{alu0101382807@ull.edu.es}
\ead{alberthf@gmail.com}
\author[b]{David D\'iaz Mart\'in}
\author[b]{Jos\'e Javier D\'iaz Garc\'ia}
\author[b]{Roger John Hoyland}
\author[b]{Luis Fernando Rodr\'iguez Ramos}
\author[b,c]{Diego Portero Rodr\'iguez}
\author[a]{Silvestre Rodr\'iguez P\'erez}
\cortext[cor1]{Corresponding author.}
\address[a]{Universidad de La Laguna (ULL), Departamento de Ingenier\'ia Industrial, ESIT, Camino San Francisco de Paula 19, San Crist\'obal de La Laguna, 38200 Santa Cruz de Tenerife, Spain}
\address[b]{Instituto de Astrof\'isica de Canarias, Electronics Department, V\'ia L\'actea s/n, San Crist\'obal de La Laguna, 38205 Santa Cruz de Tenerife, Spain}
\address[c]{Universidad de La Laguna (ULL), Departamento de Astrof\'isica, 38206 La Laguna, Tenerife, Spain}

\begin{abstract}
Frequency-multiplexed readout of kinetic inductance detector arrays measures the complex transmission of thousands of superconducting resonators. The channelizer that separates them introduces systematic effects: a gain that depends on where each tone falls in the channel grid, leakage between channels and added noise. This article presents a 4096-channel polyphase channelizer with 2/1 overlap for a 4.096~GS/s input, implemented on a first-generation radio-frequency system-on-chip, and a procedure to characterize it on the board with two receive chains, one digitizing the excitation as a reference. A split-path architecture computes the two overlap states in parallel filter branches at 256~MHz and reorders them frame by frame, under three control rules; it keeps only the transform stage at 512~MHz and affords a prototype of 16 taps per polyphase path (65\,536 coefficients) at the multiplication rate of a published 8-tap bank. Each selected channel is exported as a complex series at 1~MS/s. The measured channel response follows the coefficient-table model within 0.15~dB down to $-28$~dB, with both overlapping channels at full gain across their common band. Leakage into every measured channel 1.875~MHz or more from the tone is $-94$~dBc or lower, apart from images attributed to converter interleaving. The signal-to-noise ratio per sample lies within 1.7~dB of the limit set by the data-sheet noise density of the converter, and an emulated detector pulse is recovered with a decay time of 14.65~\us\ (standard uncertainty 0.04~\us), within 0.7\% of the expected value.
\end{abstract}


\end{frontmatter}

\section{Introduction}
\label{sec:intro}
Microwave kinetic inductance detectors (MKIDs) are superconducting pair-breaking detectors used as power detectors in millimetre and submillimetre cameras and, from the ultraviolet to the near infrared, as single-photon counters with intrinsic energy resolution~\cite{Mazin2005,Day2003,McHugh2012}. Each detector is a high-quality-factor microresonator whose resonance frequency and internal loss change when absorbed photons break Cooper pairs~\cite{Zmuidzinas2012}. Thousands of resonators can be tuned to distinct frequencies along one transmission line, so an array is read by frequency-division multiplexing: the feedline is driven with one probe tone per resonator, and the quantity to be measured, for every tone and as a function of time, is the complex transmission that its resonator imprints on it. Instruments with 2000 to 20\,000 pixels have been operated on sky in this way~\cite{Mazin2013,Meeker2018,Walter2020}. In the optical and near-infrared arrays a photon produces a phase step that rises in about a microsecond and decays in tens of microseconds~\cite{Fruitwala2020}, so the readout has to deliver each channel at a rate of the order of 1~MS/s.

The digital back end of such a readout is a measuring instrument, and its central block is the channelizer, which separates the wideband digitized signal into the resonator channels. A polyphase filter bank (PFB) followed by a fast Fourier transform (FFT) does this efficiently~\cite{Harris2003,Price2021}, and its imperfections are systematic effects on the measurand. In a critically sampled bank, a tone that falls between two channel centres is attenuated (scalloping loss): by 3.92~dB in an unwindowed FFT~\cite{Harris1978} and by 6.02~dB at the channel crossover of the prototype family used here (Section~\ref{sec:opfb}). Fabrication does not place the resonances precisely enough to align them with the channel centres~\cite{Smith2021}, so this is an amplitude error that depends on where each tone lies in the channel grid. The leakage of a tone into other channels is crosstalk between measurands, and the noise added by the arithmetic limits the resolution relative to the noise of the converter. A bank oversampled by two, with 2/1 overlap, removes the first effect~\cite{McHugh2012,Fruitwala2020,Smith2021,Yu2023}; the length of its prototype filter governs the second.

Implementing such a bank at multi-gigahertz sampling rates on a radio-frequency system-on-chip (RFSoC), which integrates the converters with the programmable logic~\cite{Smith2022,Bracken2022,Stefanazzi2022}, is constrained by timing. The closest prior design, the RFSoC bank of Smith et al.~\cite{Smith2021}, runs all its blocks at 512~MHz, with 16 parallel lanes whose vendor finite-impulse-response (FIR) cores are time-multiplexed across the two overlap states. Its prototype has 8 taps per polyphase path, reduced to 4 to close timing in the complete third-generation readout~\cite{Smith2024,SmithThesis}, and the number of taps bounds how sharp the transition between passband and stopband can be.

This article adopts the opposite trade-off, parallelism instead of clock rate, and makes three contributions. The first is a split-path architecture that computes the two overlap states in two parallel FIR branches at 256~MHz and alternates them frame by frame before the FFT, so that, within the channelizer, only the stage from the reordering memories to the FFT runs at 512~MHz. It affords a prototype of 16 taps per polyphase path, 65\,536 coefficients, with the number of real products per second of the published 8-tap bank. The second is the set of control rules that make this reordering correct under high-level synthesis (HLS), under stop and restart, and under a periodic acquisition pulse. The third is a procedure to characterize the channelizer on the board with two simultaneous receive chains, one of them digitizing the excitation as a measured reference, with its results and their dispersion for the channel response, the inter-channel isolation, the signal-to-noise ratio (SNR) and the recovery of an emulated detector pulse.

The channelizer is integrated in the embedded data acquisition system (eDAS) described by Hern\'andez et al.~\cite{Hernandez2024}, on a ZCU111 evaluation board. Hern\'andez et al.~\cite{Hernandez2026} presented the split-path topology and the frame-level reordering with their simulated channel response, and named two of the control rules; the present article states the three rules in general form and adds the integration in both receive chains with its resources, the numerical format, the characterization method with its evaluation of uncertainty and all the measurements on the board.

\section{State of the art}
\label{sec:sota}
Readout electronics for frequency-multiplexed superconducting resonators have moved from field-programmable gate array (FPGA) platforms with discrete converters, such as the ROACH boards of the CASPER collaboration~\cite{Hickish2016} used in the first two generations of optical MKID readout~\cite{McHugh2012,Fruitwala2020}, to RFSoC devices~\cite{Smith2021,Smith2024,Bracken2020,Patel2026}. Three techniques are the most common for separating the tones. An FFT synchronous with a periodic comb has no leakage between tones but yields one sample per transform, 159~Hz for up to 4000 tones in the multiplexer of van Rantwijk et al.~\cite{vanRantwijk2016}. A digital down-converter per tone centres every tone in its channel at a cost that grows with the number of tones, as in the NIKEL\_AMC boards, with up to 400 tones over 500~MHz~\cite{Bourrion2016}. A PFB delivers all channels at a rate set by its frame advance, with the prototype as the filter of every channel~\cite{Crochiere1983,Harris2003,Price2021}, and is the choice of the readouts that follow microsecond pulses~\cite{McHugh2012,Fruitwala2020,Smith2024} and of SMuRF, a tone-tracking platform for microwave resonator arrays~\cite{Yu2023}.

The prototype of a PFB can be many times longer than the transform, but when the bank is critically sampled its transition band sits at the boundary between channels. Oversampled banks move it away. In radio astronomy the first-stage channelizers of ASKAP and of the low-frequency array of the SKA oversample by 32/27, so that each channel is flat over the channel spacing and the transition falls in the overlap between neighbours~\cite{Hotan2021,Comoretto2017}; the latter uses a prototype of 14\,336 coefficients, 14 per polyphase path. In MKID readout the usual ratio has been two since the ARCONS readout~\cite{McHugh2012}, with channels twice as wide as their spacing: the second-generation optical readout uses two banks of 2048 paths and 4 taps~\cite{Fruitwala2020}, SMuRF a bank of 512 channels with 8 taps per path~\cite{Yu2023}, and the RFSoC bank of Smith et al.\ 4096 paths and 8 taps~\cite{Smith2021}. Among the reviewed readouts that publish their prototype, the banks that process a band of several gigahertz at once thus use 4 to 8 taps per polyphase path, 32\,768 coefficients at most.

What these readouts publish about the performance of their channelizers is of three kinds. The first is design figures: the prototype length, and an out-of-band rejection of about 100~dB in SMuRF~\cite{Yu2023} and of $-60$~dB as implemented ($-100$~dB in the floating-point model) in the bank of Smith et al., which was characterized as an isolated block fed with synthetic data from memory~\cite{Smith2021}. The second is measurements of the complete system, in which comb generation, analogue chain and digital processing act together: multiplexing factor, bandwidth and system noise~\cite{Bourrion2016,Patel2026}; phase-noise floors in loopback, from $-147$~dBc/Hz in the ARCONS readout~\cite{McHugh2012} to about $-80$~dBc/Hz with 2048 tones in the third-generation readout~\cite{Smith2024}; image tones of the in-phase/quadrature (I/Q) mixers, calibrated below $-30$~dBc~\cite{Fruitwala2020} or found 20 to 30~dB below the readout tones~\cite{Smith2024}; and crosstalk bounds, a requirement of $-30$~dB met in converter loopback~\cite{Smith2024} or a level dominated by the cryogenic multiplexer~\cite{Yu2023}. The third, the measured response of the channelizer itself, is scarce. Ferreyro et al.\ report an isolation between adjacent channels better than 80~dB for a Goertzel filter bank, measured with the stimulus looped back inside the FPGA, without the converters~\cite{Ferreyro2023}. Bracken et al.\ stepped the tone of a signal generator across two channels of a critically sampled bank of 256 channels and 4 taps and confirmed the loss of about 6~dB at the boundary and a leakage into the adjacent channel below their $-40$~dB requirement, with a reproducibility of about $\pm 1.1$~dB~\cite{Bracken2020}.

The converters add their own systematic effect. The 4.096~GS/s RF analogue-to-digital converters (RF-ADCs) of the first-generation RFSoC are eight-way time-interleaved; an independent characterization on the ZCU111 shows interleaving spurs at multiples of $f_s/8$ whose level depends on whether the calibration of the converter is active~\cite{Liu2021}. Offset, gain and timing mismatches between the sub-converters limit the spurious-free dynamic range of time-interleaved acquisition systems, and their estimation and adjustment remain an active subject~\cite{Yu2026,Zhao2025}. For $N$ interleaved sub-converters, gain and timing mismatches place images of an input tone of frequency $f_\mathrm{in}$ at $\pm f_\mathrm{in} + k f_s/N$, and offset mismatches place lines at $k f_s/N$~\cite{Kurosawa2001}. To our knowledge, none of the reviewed readouts reports, for an oversampled bank of thousands of channels and with RF tones at the converter inputs, the response across the boundary between two channels, the leakage into non-adjacent channels and the SNR per output sample referred to the noise of the converter. The method of Section~\ref{sec:method} measures them, with a second receive chain as reference, for a bank whose prototype is twice as long as the longest that those readouts publish.

\section{Overlapped polyphase filter bank and prototype}
\label{sec:opfb}
A filter bank with $M$ channels, a low-pass prototype filter $h[\ell]$ of length $L = KM$ and a frame advance of $D$ input samples delivers, for channel $k$ and frame $i$,
\begin{equation}
\label{eq:model}
z_k[i] = \sum_{\ell=0}^{L-1} h[\ell]\, x[iD-\ell]\, \exp\!\left[-j 2\pi k (iD-\ell)/M\right],
\end{equation}
where $x$ is the input sampled at $f_s$; the polyphase decomposition evaluates all channels with $L$ multiplications and one $M$-point FFT per frame~\cite{Crochiere1983,Harris2003}. Channel $k$ is FFT bin $k$, centred at $k f_s/M$. Equation~(\ref{eq:model}) is the measurement model of the instrument. A tone of complex amplitude $a$ at an offset $\delta f$ from the centre of channel $k$ appears in $z_k$ multiplied by the prototype response $H(\delta f)$ and rotating at $\delta f$, and in any other channel $k'$ multiplied by $H(\delta f + (k-k')\Delta f)$, where $\Delta f = f_s/M$ is the channel spacing. The departure of $|H|$ from unity over the positions that a tone can take is a systematic gain effect, and $|H|$ beyond the channel is the leakage.

Here $M = 4096$ and $f_s = 4096$~MS/s, so $\Delta f = 1$~MHz. The prototype is a Blackman-windowed sinc with $K = 16$ taps per polyphase path, $L = 65\,536$ coefficients, and cut-off $f_c = \Delta f$ (its $-6$~dB point), so each channel is 2~MHz wide between its $-6$~dB points; its linear phase gives a group delay of $(L-1)/(2 f_s) = 8$~\us. Taps are counted throughout as $K = L/M$; some authors refer the count to the decimation factor, which doubles it for a 2/1 bank~\cite{Yu2023}. Because the input is real, channels 0 to 2047 carry all the information.

\begin{figure}[tb]
\centering
\fitfigure{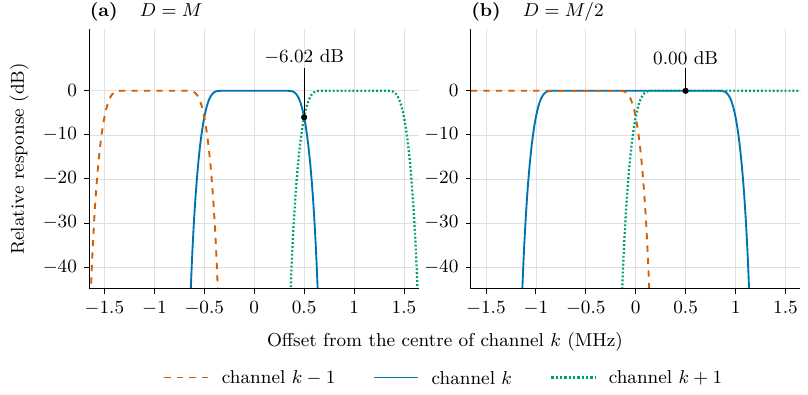}
\caption{Response of three adjacent channels in (a) a critically sampled bank (prototype cut-off $\Delta f/2$) and (b) a bank with 2/1 overlap (cut-off $\Delta f$); the dot marks the midpoint between channels $k$ and $k+1$. Blackman prototypes of 16 taps per polyphase path, $M = 4096$, $\Delta f = 1$~MHz. \aifigdata}
\label{fig:overlap}
\end{figure}

A critically sampled bank ($D = M$) cannot be flat up to the channel boundary and zero beyond it. With the same Blackman family and cut-off $\Delta f/2$, adjacent channels cross at $-6.02$~dB, as Fig.~\ref{fig:overlap}(a) shows. The overlapped bank doubles the channel width (cut-off $\Delta f$ instead of $\Delta f/2$) and halves the frame advance to $D = M/2 = 2048$, the 2/1 overlap, so that successive frames share half of their input samples. Each channel is then flat to better than 0.1~dB over $\pm 0.876$~MHz, beyond the midpoint between channels (Fig.~\ref{fig:overlap}(b)). Since any tone lies within 0.5~MHz of some channel centre, where the model response varies by less than 0.001~dB, it is received at full gain whatever its position, and every channel delivers one complex sample every $D/f_s = 0.5$~\us.

The price of the overlap is in the rotation factor of Eq.~(\ref{eq:model}), which for $D = M/2$ equals $(-1)^{ki}$ and changes the sign of the odd channels in the odd frames. Equivalently, the polyphase outputs of frame $i$ must be shifted circularly before the FFT by
\begin{equation}
\label{eq:shift}
R(i) = (i D) \bmod M,
\end{equation}
that is, by 0 or $M/2$ positions in alternate frames~\cite{Smith2021}. Realizing this shift on a 4.096~GS/s stream with 4096 channels and 16 taps is the problem that Section~\ref{sec:arch} addresses.

\begin{figure}[tb]
\centering
\fitfigure{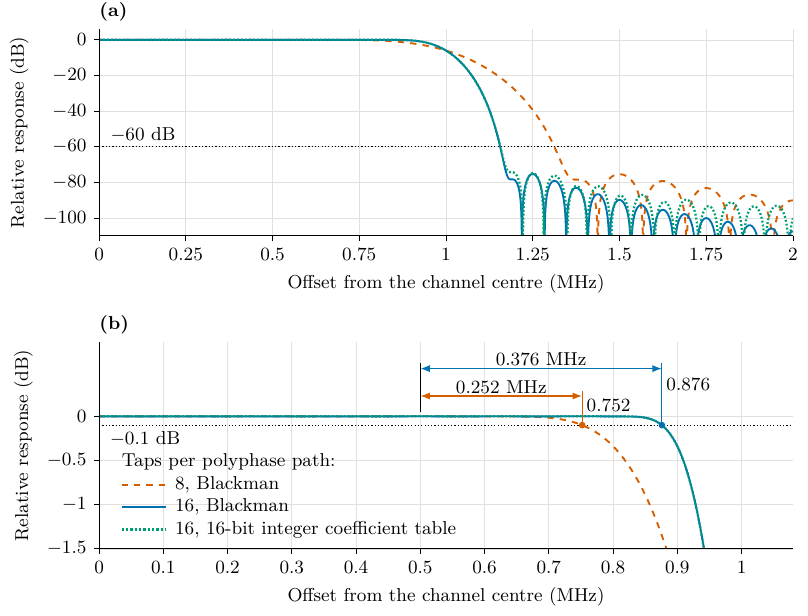}
\caption{Prototype response and filter depth. Blackman-windowed sinc prototypes with 8 and 16 taps per polyphase path ($M = 4096$, cut-off $f_c = 1$~MHz) and 16-bit integer coefficient table loaded in the hardware: (a) transition band and stopband; (b) 0.1~dB amplitude tolerance and resulting modulation margin for a tone 0.5~MHz from the channel centre. \aifigdata}
\label{fig:prototype}
\end{figure}

Channel isolation is governed by the transition band and the stopband of the prototype, and for a given window family both improve with the number of taps. Figure~\ref{fig:prototype} isolates the effect of depth within one family by comparing 8 and 16 taps per polyphase path, together with the 16-bit integer coefficient table loaded in the hardware (the prototype scaled to a peak of 32\,767 and truncated towards zero). Doubling the depth halves the transition band: the response falls from $-0.1$ to $-60$~dB in 0.28~MHz instead of 0.56~MHz, and at 1.25~MHz from the channel centre, inside the passband of the adjacent channel, it is at $-75$~dB instead of $-39$~dB. The flat region widens from 0.752 to 0.876~MHz at 0.1~dB, which for a tone 0.5~MHz from the channel centre raises the amplitude-flat margin left for its modulation sidebands from 0.252 to 0.376~MHz per side, above the $\pm 300$~kHz of flat spectrum that Smith et al.\ set as the requirement of their bank~\cite{Smith2021}. The integer table follows the ideal 16-tap response to within 0.3~dB down to $-60$~dB, and its stopband, set by the 16-bit quantization of the coefficients, stays below $-82$~dB at every offset beyond 1.34~MHz. The equiripple prototype of Smith et al.~\cite{Smith2021} is not reproduced here.

\section{Architecture}
\label{sec:arch}

\subsection{Split-path topology and clock domains}
\label{sec:split}
The converters deliver $f_s = 4096$~MS/s, which the channelizer must process in parallel words of SSR samples per clock cycle (SSR, super-sample rate), with $f_\mathrm{clk} \cdot \mathrm{SSR} = f_s$. The operating point of the filter banks is SSR${}={}$16 at 256~MHz, to which an adapter at the converter interface brings the 8 samples per 512~MHz cycle that the RF-ADC delivers.

\begin{figure}[tb]
\centering
\fitfigure{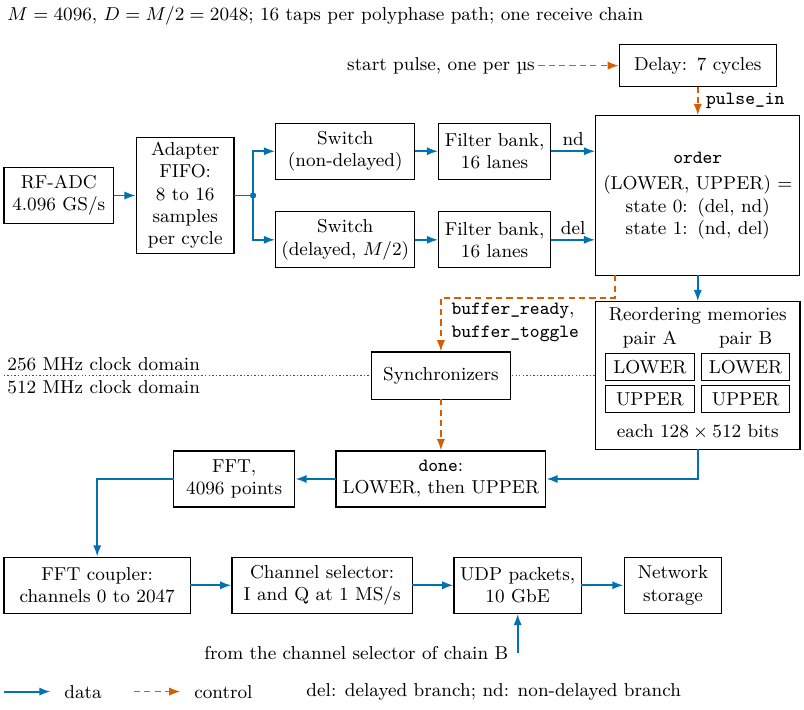}
\caption{Signal path of one channelizer in the readout bitstream, with its two clock domains. \aifigdesign}
\label{fig:system}
\end{figure}

The overlapped bank needs $M$ polyphase results for every $M/2$ new input samples, twice the output of a critically sampled bank. The design computes them in two physically parallel FIR branches, one delayed by half a frame and one non-delayed, each a 16-tap bank with 16 lanes, so that the reordering stage receives 32 results per 256~MHz cycle (Fig.~\ref{fig:system}). The second overlap state is thus produced by duplicating hardware in space instead of time-multiplexing each filter at twice the clock rate. The arithmetic rate is that of the published 8-tap bank: each channelizer computes $32 \times 256~\mathrm{MHz} \times 16 = 1.31 \times 10^{11}$ real products per second, and 16 lanes that apply 8 taps to complex samples at 512~MHz~\cite{Smith2021}, with two real products per complex tap, compute $16 \times 512~\mathrm{MHz} \times 8 \times 2$, the same number. Here the products are spent on a prototype twice as long, applied to a real input of 2.048~GHz bandwidth instead of a complex one of 4.096~GHz, with twice the clock period in every FIR stage.

Each branch has an input stage, the switch of Fig.~\ref{fig:system}, written in HLS, that implements the circular addressing of the polyphase decomposition. It buffers the input in distributed RAMs of half a frame each and presents to its filter bank, on every cycle, the 16 data words and the 16 coefficient words that the 16 lanes require. The read schedule of the delayed branch is offset by $M/2$ samples, which generates the second overlap state. Each filter bank computes the 16-tap sum of every lane as a balanced multiply--add tree on 256 DSP48E2 slices: 16-bit data (12-bit converter samples) and 16-bit coefficients give 32-bit products, and a four-level adder tree grows the word to 36 bits.

The overlapped stream that enters the FFT carries $2 f_s = 8.192 \times 10^{9}$ samples per second, so an FFT that takes 16 samples per cycle must run at 512~MHz. The two FIR branches and the reordering writer, the most resource-hungry part, work at 256~MHz; only the read side of the reordering memories, the serializer \texttt{done} and the FFT core run at 512~MHz. The data cross between the domains through the memories themselves, which are written entirely at 256~MHz and read entirely at 512~MHz, with ownership exchanged at frame boundaries; only two control bits cross (\texttt{buffer\_ready} and \texttt{buffer\_toggle}), each through a single-bit synchronizer. Downstream, a coupler retains channels 0 to 2047 of the FFT output and returns them to 256~MHz through an asynchronous first-in first-out (FIFO) memory.

\subsection{Frame-level reordering}
\label{sec:reorder}
The granularity of the reordering before the FFT is the main design decision. In the sample-level approach of Smith et al.~\cite{Smith2021,Smith2022}, implemented on the same device, each complex input sample is duplicated, the copy is delayed by half a frame and the two streams are interleaved in time, so that every lane carries both overlap states at 512~MHz through a time-multiplexed vendor FIR core. An HLS block then regroups the even and odd samples of the filtered stream and permutes them every other frame. The frame-level approach adopted here alternates the destination of whole frames instead. Each 512-bit word written to the reordering memories carries 16 samples of a single branch, and what alternates from frame to frame is the destination block, LOWER or UPPER. Because the reader always consumes a frame as the LOWER block followed by the UPPER block, exchanging which branch fills which block is the circular shift of $M/2$ positions of Eq.~(\ref{eq:shift}). Each wide word is written whole, and the permutation reduces to memory addresses and one state bit, with no selection among the 4096 positions of the frame.

\begin{figure}[tb]
\centering
\fitfigure{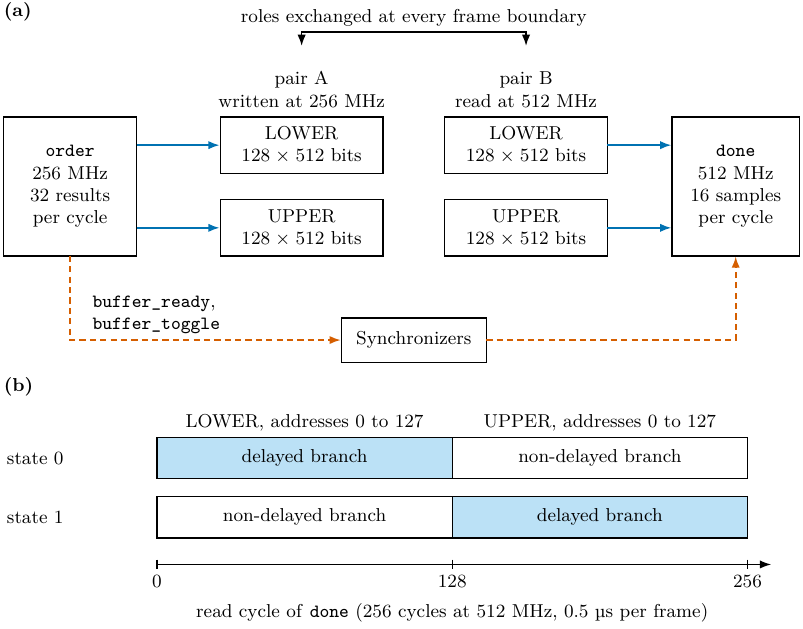}
\caption{Frame-level reordering and clock-domain crossing. (a) The two pairs of memories, one written by \texttt{order} while \texttt{done} reads the other. (b) Contents of the LOWER and UPPER blocks in the two frame states. \aifigdesign}
\label{fig:pingpong}
\end{figure}

Two modules share the mechanism (Fig.~\ref{fig:pingpong}). The writer, \texttt{order} (HLS, 256~MHz), stores the branch outputs in four block RAMs of $128 \times 512$ bits organized as two pairs, A and B, used in alternation: while one pair is written, the other is read. The reader, \texttt{done} (VHDL, 512~MHz), takes the released pair, its 128 LOWER words and then its 128 UPPER words, 256 read beats in the same 0.5~\us\ as the 128 writes, and delivers 16 samples per cycle to the FFT. The writer keeps three state variables: a write pointer over the 128 addresses of the active pair, a pair selector that changes at each frame boundary, and a frame-state bit that alternates the LOWER and UPPER destinations of the two branches. When packing, \texttt{order} keeps bits 35 to 4 of each 36-bit filter sample, so that 16 samples fill one 512-bit word.

\subsection{Control rules}
\label{sec:rules}
Three rules make the reordering correct. Hern\'andez et al.~\cite{Hernandez2026} name the first two; to our knowledge the readout literature does not otherwise document them, and each is stated here in general form.

The first rule concerns the joint update of the state. At the end of each frame the pair selector and the frame-state bit must change together. The HLS tool is free to schedule independent operations into different pipeline stages even when they are adjacent in the source; with the two toggles one stage apart the design would see, for one cycle per frame, the new pair with the old overlap state. Such a fault leaves the data flowing, raises no simulation error and shows only in the spectral response. All frame-boundary toggles are therefore grouped into a single update expression, which the tool schedules in one stage. In general, state variables whose joint value defines correctness must be updated through a single expression.

The second rule concerns stop and restart, since acquisitions are started and stopped from the host. The writer has a frame-valid level, de-asserted while the acquisition is stopped, and a synchronization input (\texttt{pulse\_in}) that returns the write pointer and the two state bits to a known phase in one cycle and withdraws the buffer-ready flag seen by the reader. A full reset on every stop would restart the permutation at an arbitrary phase with respect to the circular buffers of the input stages. On stop, therefore, only the flag that gates the first output frame is cleared, which withdraws the buffer-ready flag and returns the reader to idle once the frame in progress is complete, while the write pointer, the pair selector and the frame-state bit are preserved. In general, only the state that would break the alignment if it were left stale is reset on stop, and the counters and pointers that encode the phase between pipeline stages are preserved.

The third rule concerns the alignment of a periodic pulse. In the integrated system an acquisition trigger in the 256~MHz domain issues one start pulse every 256 cycles (1~\us) to each chain, while the filter branches mark one frame every 128 cycles (0.5~\us), and \texttt{order} receives the pulse on its synchronization input, so that its frame phase is re-established every microsecond. The frame boundary of the filter banks reaches \texttt{order} nine cycles after the start pulse, whereas \texttt{order} resumes writing two cycles after its synchronization input. The pulse is therefore delayed by seven cycles (27.3~ns), so that every frame written to the memories holds the samples of a single filter-bank frame under a single permutation state. In general, a periodic synchronization of a reordering stage must be referred to the frame boundary at its own input.

\subsection{Exported series and numerical format}
\label{sec:format}
Because the pulse arrives once every two filter-bank frames, in each microsecond \texttt{order} completes one frame, \texttt{done} delivers it to the FFT, and the next pulse returns the write pointer to the frame boundary before the second pair is complete. The integrated system thus exports every second frame: each exported channel is a complex series at $f_\mathrm{exp} = (f_s/D)/2 = 1$~MS/s, while every internal stage remains sized for one frame every 0.5~\us. This rate matches one resonator tone per channel. The tone is read in the nearest channel, at most 0.5~MHz from its centre; with its modulation inside the flat margin of $\pm 0.376$~MHz, the occupied band is at most 0.752~MHz wide and fits in the 1~MHz interval of the exported series wherever the tone lies. The export rate does not change the flat gain provided by the overlap, which is a property of the channel response. The rate is also that of the second-generation optical readout, which samples each channel every 1.024~\us~\cite{Fruitwala2020}.

The word format follows from the growth that the filter can reach. A 12-bit sample does not exceed $2^{11}$ in magnitude, and the largest sum of the absolute values of the 16 integer coefficients of a polyphase path is 42\,221, so the output $S$ of every path satisfies
\begin{equation}
\label{eq:window}
|S| \le 2^{11} \times 42\,221 = 86\,468\,608 < 2^{27} .
\end{equation}
Bits 35 to 27 of $S$ therefore repeat its sign, and \texttt{done} passes bits 27 to 10 of the filter output to the 18-bit FFT input, without rounding and without loss of range: a full-scale tone does not exceed about 65\,500 counts, within the range $|x| < 2^{16}$ in which the FFT core excludes internal overflow. Taking instead the 18 most significant bits would spend eight of them on the sign; a tone at $-1$~dBFS would then occupy about 58 counts at the output, and the SNR per output sample would fall, in the simulation of Section~\ref{sec:simres}, from 84.9 to 41.3~dB. The FFT, with the full $2^{-12}$ scaling, returns 18-bit real and imaginary components. For each chain a selector forwards the two components of one channel to the packet transport, which groups 2048 samples of each component in one User Datagram Protocol (UDP) packet of the 10~Gigabit Ethernet (10~GbE) link.

\section{Implementation}
\label{sec:impl}
The design is built with Vivado and Vitis HLS 2022.1 for the XCZU28DR device (speed grade $-2$) of the ZCU111, with the FFT core generated in Vitis Model Composer. Table~\ref{tab:impl} gives the post-implementation figures of the complete bitstream, which contains two channelizers (one per receive chain), a waveform-playback path for the digital-to-analogue converters, the 10~GbE transport and the processor interconnect.

\begin{table}[tb]
\centering
\caption{Post-implementation utilization, timing and power of the complete bitstream on the XCZU28DR. The measured power is the sum of the 14 power monitors of the board during the measurements.}
\label{tab:impl}
\footnotesize
\begin{tabular}{@{}lll@{}}
\toprule
Resource & Used & Device or condition \\
\midrule
LUTs & 275\,244 (64.7\%) & 425\,280 \\
\quad of which as memory & 96\,453 (45.2\%) & 213\,600 \\
Registers & 214\,223 (25.2\%) & 850\,560 \\
Block RAM (36~kbit blocks) & 365 (33.8\%) & 1080 \\
DSP48E2 slices & 1312 (30.7\%) & 4272 \\
\quad of which in the filter banks & 1024 & $4 \times 256$ \\
Set-up slack (512 / 256~MHz) & $0.000$ / $+0.060$~ns & all constraints met \\
Hold / pulse-width slack & $+0.006$ / $+0.030$~ns & \\
Power, measured & 12.89~W (8.03~W at 0.85~V) & logic at 45.3~$^\circ$C \\
\bottomrule
\end{tabular}
\end{table}

The dominant arithmetic cost is the pair of filter banks of each channelizer, 512 DSP48E2 slices (12\% of the device), the direct price of computing its real products at 256~MHz and not at 512~MHz. The polyphase buffers are held in distributed RAM, which leaves the block-RAM usage at the 33.8\% of Table~\ref{tab:impl} at the cost of look-up tables (LUTs) configured as memory.

Timing is met on all constraints over $1.36 \times 10^{6}$ endpoints. The worst set-up path lies inside the FFT core at 512~MHz, with zero slack and 82.5\% of its delay in routing, while the 256~MHz domain, which holds the four filter banks, closes with a positive margin.

\section{Characterization method}
\label{sec:method}

\subsection{Measurement set-up}
\label{sec:setup}
\begin{figure}[tb]
\centering
\fitfigure{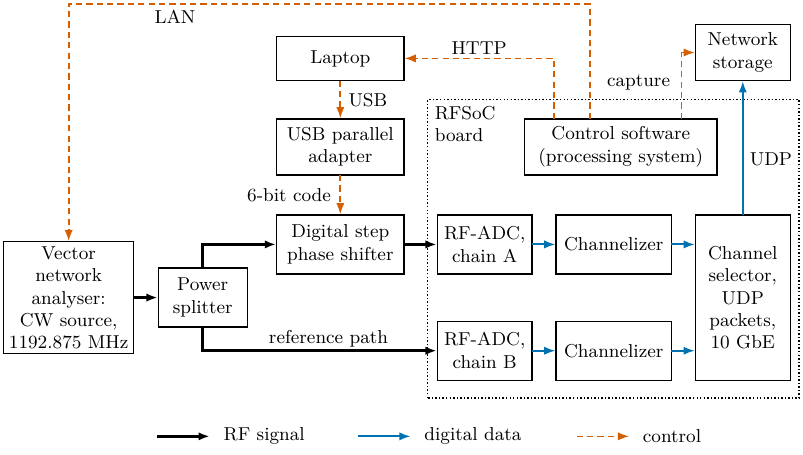}
\caption{Measurement set-up, with chain~A fed through the digital step phase shifter and chain~B by the reference path. CW: continuous-wave (unmodulated) tone. \aifigdesign}
\label{fig:setup}
\end{figure}

Figure~\ref{fig:setup} shows the set-up. A vector network analyser (Rohde \& Schwarz ZNB20) provides an unmodulated tone, which a power splitter divides into two paths. The path that feeds chain~A passes through a digital step phase shifter (pSemi PE44820~\cite{PE44820DS}) and enters one RF-ADC of the board; the other path enters a second RF-ADC directly and feeds chain~B, which digitizes the excitation itself as a reference. Comparing two converter channels driven by one split signal, a zero measurement, has been used on this evaluation board to determine the noise floor of a phasemeter~\cite{Subrahmanya2025}; here each chain has its own channelizer. Both converters sample the RF signal directly in the first Nyquist zone, without analogue down-conversion, and each chain exports one selected channel. Unless stated otherwise the tone is at 1192.875~MHz, 125~kHz below the centre of channel 1193, at $-10$~dBm at the analyser output. All the results come from one bitstream in a single session, in which the 154 captures behind the figures and tables amount to 635\,132 packets of 8214~bytes of UDP payload, with none dropped at the receiver.

\subsection{Measurands and estimators}
\label{sec:estim}
Metrological terms are used as defined in the International Vocabulary of Metrology~\cite{JCGM2012}. In this characterization the measurand is the complex amplitude of a tone in a channel, $z[m]$, in counts of the 18-bit output, with $m$ the index of the exported sample. The channel response for a tone at a given offset is estimated as the mean modulus of the samples of a capture, $\langle |z[m]| \rangle$, normalized to its value with the tone at the channel centre.

With the tone in channel $k_\mathrm{t}$, the complex leakage into channel $k_\mathrm{t}+n$ is estimated by coherent cross-correlation,
\begin{equation}
\label{eq:leak}
c_n = \frac{\langle z_{k_\mathrm{t}+n}[m]\, z_{k_\mathrm{t}}^{*}[m] \rangle}{\langle |z_{k_\mathrm{t}}[m]|^{2} \rangle},
\end{equation}
and expressed as $20 \log_{10} |c_n|$ in dBc. The average accumulates the component of channel $k_\mathrm{t}+n$ that rotates with the tone, which is what channelizer leakage produces, and averages out the noise and any component that rotates in the opposite sense. On the board, chain~A exports channel $k_\mathrm{t}$ and chain~B channel $k_\mathrm{t}+n$, and $c_n$ is divided by the complex gain ratio between the two chains, measured with both on channel $k_\mathrm{t}$.

The SNR per output sample is
\begin{equation}
\label{eq:snr}
\mathrm{SNR} = 10 \log_{10} \frac{A^{2}}{2\sigma^{2}},
\end{equation}
with $A$ the amplitude of the tone and $\sigma$ the standard deviation of the noise per component. $A$ is the square root of the power in the main lobe of the windowed periodogram of the tone, averaged over the packets, and $\sigma$ is measured with the RF output of the analyser off, so that the SNR characterizes the additive noise of the receive chain separately from the noise of the source. The reference is the noise of the converter: $-151$~dBFS/Hz is the value that the data sheet gives as typical with a 1.9~GHz input at $-1$~dBFS and as maximum with the input at the noise floor~\cite{DS926}. Over the noise-equivalent bandwidth of the channel, $f_s \sum_\ell h^2[\ell] / \bigl(\sum_\ell h[\ell]\bigr)^2 = 1.94$~MHz, this density sets a limit of 87.1~dB per sample for a tone at $-1$~dBFS, whose amplitude is 14\,604 counts in the simulation of Section~\ref{sec:simref}; the input level referred to the converter is therefore $P = -1~\mathrm{dBFS} + 20 \log_{10}(A/14\,604)$, and the limit at that level, $87.1~\mathrm{dB} + (P + 1~\mathrm{dB})$.

The phase of chain~A is read relative to the reference chain,
\begin{equation}
\label{eq:relphase}
\varphi_\mathrm{AB}[m] = \arg\!\left( z_\mathrm{A}[m]\, z_\mathrm{B}^{*}[m] \right),
\end{equation}
which cancels the phase of the analyser and of the sampling clock, common to both chains. With $\sigma \ll A$ the noise of each chain perturbs its phase by $\sigma/A$ radians, so the standard deviation of the relative phase per sample is
\begin{equation}
\label{eq:phasenoise}
\sigma_\varphi \simeq \sqrt{ \left( \sigma_\mathrm{A}/A_\mathrm{A} \right)^{2} + \left( \sigma_\mathrm{B}/A_\mathrm{B} \right)^{2} } .
\end{equation}

The sample period is obtained with both chains reading the same tone in channels 1193 and 1194, whose centres are $\Delta f = 1$~MHz apart by construction. If $T$ is the period between consecutive samples of a packet, the phase advances per sample of the two channels differ, in cycles, by $\Delta\nu = \Delta f\, T - 1$, which does not depend on the frequency of the tone and refers $T$ to the sampling clock and not to the analyser.

\subsection{Evaluation of uncertainty}
\label{sec:unc}
The uncertainties quoted with the results are Type~A evaluations in the sense of the Guide to the Expression of Uncertainty in Measurement~\cite{JCGM2008}, obtained from the dispersion of repeated observations and given as standard uncertainties unless an interval is stated. The sample period is bounded by the range of $\Delta\nu$ over 15 tone positions. The repeatability under restart is the range of each quantity over 20 acquisitions. For the leakage, each capture of about 1.6 million samples is divided into 20 segments, and the standard uncertainty of $c_n$ is the standard deviation of the 20 segment estimates divided by $\sqrt{20}$; it lies between $-119$ and $-126$~dBc, and estimates close to it are read as upper bounds. The noise $\sigma$ is the experimental standard deviation of each component over 2.8 million complex samples per chain. The decay time of the emulated pulse is fitted to the complex mean of the pulses of a record, and its standard uncertainty and its 95\% coverage interval come from a bootstrap with 300 resamples of those pulses. The code-to-phase law of the phase shifter is measured on the same set-up, with a median dispersion of $0.08^\circ$ between repetitions over its 63 codes. Two systematic effects are kept apart from these figures: the input level $P$ is referred to the converter through the simulated scale, so the distance between the measured SNR and the limit depends only on $\sigma$, and the tone images of the interleaved converter (Section~\ref{sec:isolation}) are reported separately from the leakage of the channelizer.

\subsection{Simulation reference}
\label{sec:simref}
The measurements are compared with two kinds of simulation in XSim, the simulator of the design suite. A batch of 100 simulations of the filter banks and of the register-transfer-level (RTL) reordering modules, each with a single tone between 20.0 and 29.9~MHz in steps of 100~kHz, gives the response of channels 20 to 28 in their passbands and at their crossovers, each channel being normalized to its maximum over the sweep. The complete integrated design, from the RF-ADC interface to the output of the FFT coupler, is simulated with the RTL sources and the intellectual-property cores of the implemented bitstream; the stimulus is a tone plus white Gaussian noise at the data-sheet noise density of the converter ($-151$~dBFS/Hz), quantized to 12 bits at 4.096~GS/s.

\section{Results}
\label{sec:results}

\subsection{Simulation results}
\label{sec:simres}
In the sweep, a tone at the centre of a channel is recovered there at the reference level, and the two adjacent channels respond at $-6.02$~dB, the crossover of the overlapped passbands for a tone 1~MHz from their centres, symmetrically to within 0.01~dB. Within $\pm 0.8$~MHz of every channel centre the amplitude deviates from the reference by less than 0.005~dB on the 100~kHz grid. The symmetry is the sensitive figure, because the faults that the first two rules of Section~\ref{sec:rules} prevent appear as asymmetric adjacent-channel responses or as frame-to-frame instability of the amplitude.

In the integrated design, a tone at $-1$~dBFS at the centre of a channel gives an SNR of 84.9~dB per output sample, 2.2~dB below the 87.1~dB that the noise of the stimulus alone would leave in the channel without further quantization; with the acquisition trigger of the board the result is 84.8~dB. With the tone 125~kHz below the centre of channel $k_\mathrm{t}$, channel $k_\mathrm{t}-1$ receives it at $-0.1$~dB, which is the coverage of the overlap and not leakage. Channel $k_\mathrm{t}+1$, whose centre lies 1.125~MHz from the tone, inside the transition band, receives it at $-39.4$~dBc, as the coefficient table predicts. The 27 channels examined 1.875~MHz or more away, from $n = -6$ to $n = +1000$, lie between $-85.0$ and $-97.9$~dBc with the 11 frames that the simulation averages.

\subsection{Output cadence and restart repeatability}
\label{sec:cadence}
Over the 15 tone positions of the sweep of Section~\ref{sec:response} in which channels 1193 and 1194 both receive the tone at full gain, $\Delta\nu$ lies between $-1.2 \times 10^{-6}$ and $+1.6 \times 10^{-6}$, with a mean of $0.81 \times 10^{-6}$: the period between consecutive samples of a packet is 1~\us\ to within 2 parts per million of the sampling clock, as designed. Twenty acquisitions were started and stopped from the host about 15~s apart, without reloading the bitstream and with the tone at $+10$~dBm. Over the twenty, the amplitude of each chain repeats to within 0.07\% and the phase of chain~A relative to chain~B to within $0.016^\circ$ (from $42.811^\circ$ to $42.827^\circ$), where a slip of one sample between the chains would shift it by $45^\circ$: the frame alignment is recovered at every restart.

\subsection{Channel response}
\label{sec:response}
\begin{figure}[tb]
\centering
\fitfigure{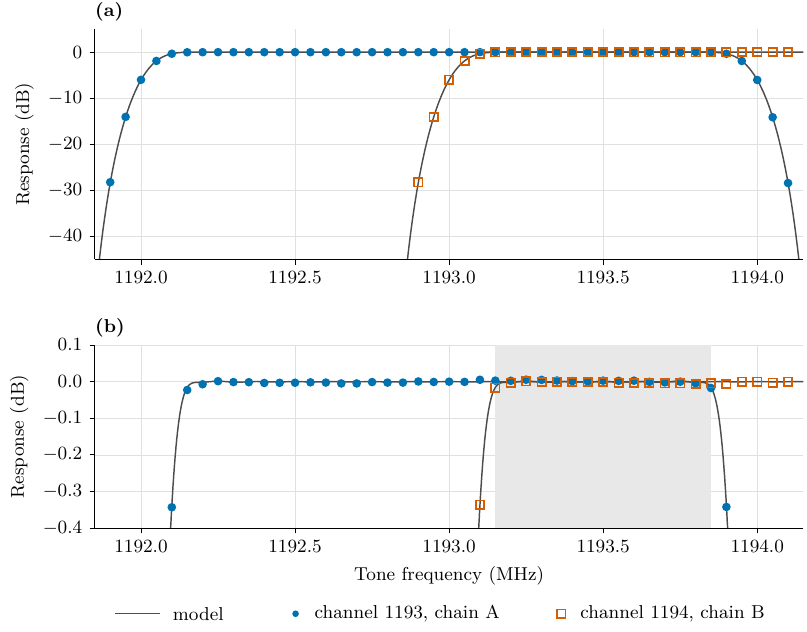}
\caption{Channel response measured on the board. Mean modulus of channels 1193 (chain~A) and 1194 (chain~B) for 45 tones 50~kHz apart at $-10$~dBm, normalized at each channel centre, and model of the integer coefficient table (line): (a) passbands and transition bands; (b) detail of the passbands, with both channels at full gain across the shaded band. \aifigdata}
\label{fig:response}
\end{figure}

A sweep of 45 tones from 1191.9 to 1194.1~MHz in steps of 50~kHz, with chain~A exporting channel 1193 and chain~B channel 1194, maps two adjacent channels (Fig.~\ref{fig:response}). The measured response follows the model of the integer coefficient table point by point: at 0.95, 1.00, 1.05 and 1.10~MHz from the centre of channel 1193 it is $-1.91$, $-6.00$, $-14.05$ and $-28.23$~dB on the lower flank and $-1.92$, $-6.04$, $-14.13$ and $-28.41$~dB on the upper flank, against $-1.91$, $-6.02$, $-14.10$ and $-28.36$~dB in the model. The largest deviation from the model is 0.05~dB down to $-14$~dB and 0.13~dB at $-28$~dB, with opposite signs on the two flanks: the rotation of the tone in the channel places it about 225~Hz above its nominal frequency, the difference between the frequency reference of the analyser and the clock of the board, and with the model evaluated at that frequency the largest deviation is 0.05~dB. Within $\pm 0.85$~MHz of the centre the response varies by 0.028~dB peak to peak in channel 1193 and by 0.017~dB over the measured part of channel 1194, against 0.017~dB in the model, and within $\pm 0.8$~MHz every point lies within 0.01~dB of the centre value. Over 1193.15--1193.85~MHz both channels receive the tone at full gain, which is the direct observation, with the real converter, that the overlap removes the scalloping loss at the channel boundary.

\subsection{Inter-channel isolation}
\label{sec:isolation}
\begin{figure}[tb]
\centering
\fitfigure{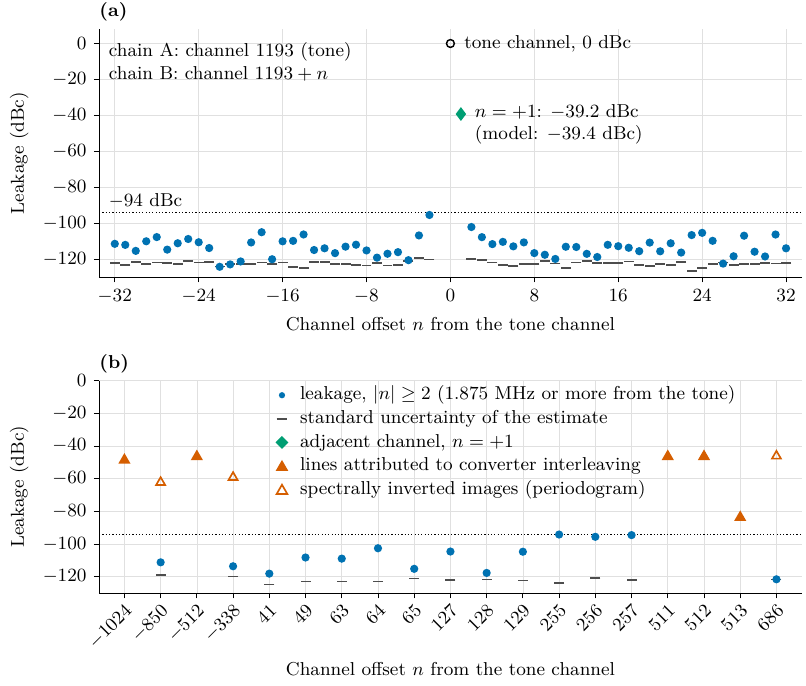}
\caption{Inter-channel isolation measured on the board. Leakage $20 \log_{10} |c_n|$, Eq.~(\ref{eq:leak}), from the tone channel 1193 into channel $1193 + n$, with the tone 125~kHz below the channel centre: (a) the tone channel, $n = +1$ and $2 \le |n| \le 32$; (b) distant channels and lines attributed to converter interleaving (filled triangles: estimate of Eq.~(\ref{eq:leak}); open triangles: spectrally inverted images, from the averaged periodogram of the channel). Short dashes: standard uncertainty of each estimate; dotted line: $-94$~dBc. \aifigdata}
\label{fig:isolation}
\end{figure}

With chain~A on the tone channel $k_\mathrm{t} = 1193$ and chain~B on channel $k_\mathrm{t}+n$, the leakage over $2 \le |n| \le 32$ and in 14 more distant channels, up to 850 positions away, lies between $-94.2$ and $-124.2$~dBc (Fig.~\ref{fig:isolation}); the values below about $-115$~dBc are close to the standard uncertainty of the estimator and are read as upper bounds. In the nearest of these channels, $n = -2$, it is $-95.4$~dBc, against $-95.3$~dB in the model of the coefficient table. With the tone at the centre of its channel the leakage into $n = 2$ to 5 lies between $-99.6$ and $-119.4$~dBc. Channel $k_\mathrm{t}+1$ receives $-39.2$~dBc, against $-39.4$~dBc in the simulation and in the model. The only channels in which the estimate is higher are those that contain the images of the tone at $f_\mathrm{in} - f_s/8$, $f_\mathrm{in} + f_s/8$ and $f_\mathrm{in} - f_s/4$ ($n = -512$, $+512$ and $-1024$), at $-46$ to $-49$~dBc, together with channel $n = +511$, which receives the $+512$ image through its overlapping passband ($-46.5$~dBc), and channel $n = +513$, which receives it through its transition band ($-83.7$~dBc). We attribute these lines to the converter and not to the channelizer. They fall where a time-interleaved converter places images of a tone~\cite{Jenq1988}: at $\pm f_\mathrm{in} + k f_s/8$ for the gain and timing mismatches of an eight-way converter (Section~\ref{sec:sota}), as the documentation of the device also states for its RF-ADCs~\cite{PG269,DS926}. Of the seven such positions for this tone, six are among the measured channels. In the three with lines the image keeps the sense of rotation of the tone. In the other three ($n = -850$, $-338$ and $+686$) it is spectrally inverted: Eq.~(\ref{eq:leak}) averages it out, with estimates of $-111$~dBc or lower, and the periodogram of the channel, averaged over the packets, shows it at $-62$, $-59$ and $-46$~dBc. In distant channels without an image position the periodogram shows no line above its floor of about $-82$~dBc. The level of the lines differs between the two RF-ADCs, drifts in time with the same configuration and changes when the background-calibration mode of the converter is switched (in channel 681, from $-46.6$ to $-52.6$~dBc in one converter and from $-45.0$ to $-52.3$~dBc in the other), whereas the processing of the channelizer is deterministic and identical in both chains.

\subsection{Noise and signal-to-noise ratio}
\label{sec:snr}
With the RF output off, the noise of the two chains is $\sigma = 0.555$ and 0.553~counts (counts of the 18-bit output) per component, against 0.59~counts in the simulation. The mean of the samples, $-0.53$~counts in the real component and $-0.24$~counts in the imaginary one, is the bias of an arithmetic that truncates without rounding and agrees between the chains to 0.001~counts. Table~\ref{tab:snr} gives the SNR at three levels of the analyser. It grows 10~dB per 10~dB of input in both chains, and at every level it lies 1.7~dB below the limit set by the data-sheet noise density of the converter ($-151$~dBFS/Hz). The distance, the same at every level by construction (Section~\ref{sec:unc}), is smaller than the 2.2~dB of the simulation because the noise of these converters is slightly below that of the simulated stimulus. Chain~A lies 4.5~dB below chain~B, the insertion loss of the phase shifter.

\begin{table}[tb]
\centering
\caption{Signal-to-noise ratio per output sample measured on the board, Eq.~(\ref{eq:snr}), with the tone at 1192.875~MHz in channel 1193 of both chains. Output: level at the analyser output; $P$, $A$ and the limit as defined in Section~\ref{sec:estim}.}
\label{tab:snr}
\footnotesize
\begin{tabular}{@{}cccccc@{}}
\toprule
Output (dBm) & Chain & $P$ (dBFS) & $A$ (counts) & SNR (dB) & Limit (dB) \\
\midrule
$-10$ & A & $-30.8$ & 472.9 & 55.6 & 57.3 \\
 & B & $-26.3$ & 794.7 & 60.1 & 61.8 \\
0 & A & $-20.8$ & 1496.2 & 65.6 & 67.3 \\
 & B & $-16.3$ & 2514.4 & 70.1 & 71.8 \\
$+10$ & A & $-10.7$ & 4784.6 & 75.7 & 77.4 \\
 & B & $-6.2$ & 8031.8 & 80.2 & 81.9 \\
\bottomrule
\end{tabular}
\end{table}

\subsection{Recovery of an emulated detector pulse}
\label{sec:pulse}
\begin{figure}[tb]
\centering
\fitfigure{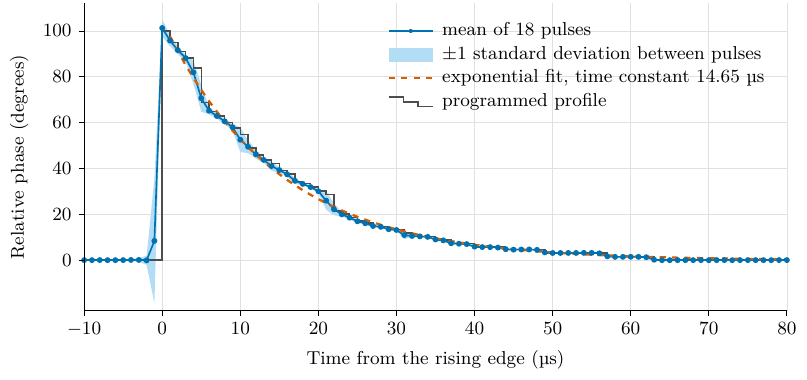}
\caption{Emulated detector pulse read at 1~MS/s. Phase of chain~A relative to chain~B, Eq.~(\ref{eq:relphase}), while the phase shifter steps one code per microsecond; the programmed profile is converted to phase with the measured law. \aifigdata}
\label{fig:pulse}
\end{figure}

To emulate the phase response of a detector to a photon, the phase shifter in the path of chain~A is driven in its direct parallel mode by a USB interface that applies a new 6-bit code every microsecond, from code 63 to 0 along an exponential with a programmed time constant of 15~\us. The code-to-phase law measured on the set-up gives $99.93^\circ$ for code 63, an average of $1.59^\circ$ per code. At 1.19~GHz, below the specified band of the device, its bits are not exactly binary-weighted, and the transitions 48$\to$47, 32$\to$31 and 16$\to$15 produce steps of 13.0, 4.4 and $7.1^\circ$; the expected response is therefore computed from the measured law. Figure~\ref{fig:pulse} shows the complex mean of 18 pulses of one record, aligned at the 50\% crossing of the step. The step of $101^\circ$ rises from 10\% to 90\% in 0.87~\us, less than one sample period, the time resolution of the exported series. The decay is fitted with a time constant $\tau = 14.65$~\us, with a standard uncertainty of 0.04~\us\ and a 95\% coverage interval from 14.55 to 14.70~\us, 0.7\% below the 14.75~\us\ expected. Sending the same code sequence with 2 and 5~\us\ per code gives $\tau = 29.50$ and 75.80~\us\ (standard uncertainties 0.04 and 0.06~\us), against 29.70 and 76.20~\us\ expected: in all three cases the fitted decay lies within 1\% of the expected one. The noise of the relative phase on the baseline is $0.085^\circ$ (1.5~mrad) per sample, close to the $0.078^\circ$ that Eq.~(\ref{eq:phasenoise}) gives for the additive noise of the two chains at this level (Table~\ref{tab:snr}).

\subsection{Operation with a detector array}
\label{sec:array}
The readout system, configured with a critically sampled filter bank, has been operated with an MKID array in a cryostat, where it identified the resonances of individual pixels in the dark~\cite{Hernandez2024,Hernandez2026}. The array, with resonances between 0.9 and 1.8~GHz, within the first Nyquist zone of the converters, was held at about 100~mK in an adiabatic demagnetization refrigerator, behind 80~dB of attenuation and ahead of a low-noise amplifier at 4~K.

\section{Discussion}
\label{sec:discussion}
Filter banks oversampled by two are established practice in MKID readout~\cite{McHugh2012,Fruitwala2020,Yu2023}; this work contributes a way of realizing the oversampling, the prototype that it affords and a measured characterization of the result. Table~\ref{tab:comparison} sets the design against the closest one, the bank of Smith et al.~\cite{Smith2021}, which implements the same 2/1 function on the same device with the same channel count and width.

\begin{table}[tb]
\centering
\caption{Comparison between the sample-level bank of Smith et al.~\cite{Smith2021,Smith2024,SmithThesis} and the frame-level design of this work. Both have $M = 4096$ channels of 2~MHz spaced 1~MHz apart and take a 4.096~GS/s input.}
\label{tab:comparison}
\footnotesize
\begin{tabular}{@{}p{0.19\textwidth}p{0.37\textwidth}p{0.37\textwidth}@{}}
\toprule
Aspect & Smith et al. & This work \\
\midrule
Input & Complex, two converters (I and Q) & Real, one RF-ADC per channelizer \\
Prototype & 32\,768 coefficients, equiripple, 8 taps (4 in the complete readout) & 65\,536 coefficients, Blackman, 16 taps \\
Overlap states & Time-multiplexed in 16 vendor FIR cores; sample-level reordering & Two custom HLS FIR branches; frame-level reordering \\
Clocking & Whole bank at 512~MHz & 256~MHz; reordering memories, \texttt{done} and FFT at 512~MHz \\
Real products per second & $1.31 \times 10^{11}$ (8-tap bank) & $1.31 \times 10^{11}$ per channelizer \\
Reported rejection & $-60$~dB of image rejection as implemented ($-100$~dB in floating point) & Coefficient table below $-82$~dB beyond 1.34~MHz; $-94$~dBc or lower measured 1.875~MHz or more from the tone \\
Reported scope & 8 taps: isolated bank fed with synthetic data; 4 taps: complete readout & 16 taps: two channelizers in a complete design fed by RF-ADCs, measured on the board \\
\bottomrule
\end{tabular}
\end{table}

The two designs make opposite trade-offs. The sample-level design keeps the number of multipliers low at the price of half the clock period in every stage (Table~\ref{tab:comparison}). The frame-level design duplicates the filter in space, 512 DSP slices per channelizer, and spends the same number of real products per second on a prototype twice as long. That prototype, 65\,536 coefficients, is to our knowledge the longest among the implemented analysis filter banks of the reviewed readouts that publish it; longer ones have been evaluated only in simulation~\cite{Bracken2020}. Among the reviewed readouts that describe their oversampling scheme, the design is also the only one in which two spatial FIR branches share one FFT through frame-level reordering.

The rejection figures of Table~\ref{tab:comparison} are obtained differently, those of the sample-level bank from synthetic data fed to an isolated block~\cite{Smith2021} and those of this work from RF tones at the converters of the complete design, and are not claimed to be equivalent. The same holds for the isolation better than 80~dB that Ferreyro et al.\ measure inside the FPGA, without the converters~\cite{Ferreyro2023}.

Which trade-off is preferable depends on the array. For sparse arrays the compact design wins on resources. For dense arrays, where the leakage between adjacent resonators is the limiting systematic effect, the longer prototype converts DSP slices into isolation: compared with 8 taps of the same family it gives about 36~dB more rejection 1.25~MHz from the channel centre and an amplitude-flat margin of $\pm 0.376$~MHz for a tone at any position. In the design measured here the channelizer is no longer what limits the measurement: the lines that remain are attributed to the converter, and the SNR per sample is set by the converter noise to within 1.7~dB.

The characterization method does not depend on the architecture. It needs two receive chains and a splitter and uses the excitation itself as the reference. It reaches a repeatability of 0.07\% in amplitude over 20 restarts and, with captures of about 1.6 million samples, a standard uncertainty of the leakage estimate between $-119$ and $-126$~dBc, which is what makes a leakage of $-94$~dBc measurable through the converters and distinguishes the converter lines from those of the channelizer. System-level figures such as multiplexing factor and readout noise~\cite{Bourrion2016,Patel2026} do not give this information.

\section{Conclusions}
\label{sec:conclusions}
A polyphase channelizer of 4096 channels with 2/1 overlap has been presented together with a procedure to characterize it as a measuring instrument. Computing the two overlap states in parallel FIR branches at 256~MHz and reordering whole frames confines the 512~MHz clock of the channelizer to the stage between the reordering memories and the FFT. The number of real products per second of the published 8-tap bank with the same channel count on the same device is spent on a prototype of 16 taps per polyphase path, which halves the transition band compared with 8 taps of the same family. Three control rules (joint update of the state, selective reset on stop, and alignment of a periodic pulse with the frame boundary) make the design correct and restartable. Two channelizers close timing in a complete readout bitstream on a first-generation RFSoC, and each selected channel is exported as a complex series at 1~MS/s.

Measured on the board with two simultaneous chains, one of them a reference, the channel response follows the coefficient-table model within 0.15~dB down to $-28$~dB, with both overlapping channels at full gain across their common band. The leakage is $-94$~dBc or lower in every measured channel 1.875~MHz or more from the tone, the only stronger lines being the images attributed to converter interleaving. The SNR per output sample lies within 1.7~dB of the limit set by the data-sheet noise density of the converter ($-151$~dBFS/Hz). Amplitude and relative phase repeat to within 0.07\% and $0.016^\circ$ over 20 restarts, and an emulated detector pulse is recovered with a decay time within 0.7\% of the expected value and a phase noise of $0.085^\circ$ per sample. The next step is to read a detector array in the cryostat with this bitstream; work continues on the generation of the excitation comb and on real-time phase processing for single-photon detection.

\section*{CRediT authorship contribution statement}
\textbf{Alberto Hern\'andez Fern\'andez:} Conceptualization, Methodology, Software, Validation, Formal analysis, Investigation, Data curation, Visualization, Writing -- original draft, Writing -- review \& editing. \textbf{David D\'iaz Mart\'in:} Conceptualization, Methodology, Software, Validation, Investigation, Supervision, Writing -- review \& editing. \textbf{Jos\'e Javier D\'iaz Garc\'ia:} Supervision, Project administration, Resources, Writing -- review \& editing. \textbf{Roger John Hoyland:} Supervision, Resources, Writing -- review \& editing. \textbf{Luis Fernando Rodr\'iguez Ramos:} Supervision, Project administration, Resources, Writing -- review \& editing. \textbf{Diego Portero Rodr\'iguez:} Investigation, Writing -- review \& editing. \textbf{Silvestre Rodr\'iguez P\'erez:} Supervision, Writing -- review \& editing.

\section*{Funding}
This research did not receive any specific grant from funding agencies in the public, commercial, or not-for-profit sectors.

\section*{Acknowledgements}
The authors thank the Instrumentation Area of the Instituto de Astrof\'isica de Canarias (IAC) and the Universidad de La Laguna (ULL) for their support in the development of the eDAS. D.P.R. is ``personal contratado predoctoral del Programa de Astrof\'isicos Residentes del IAC''.

\section*{Declaration of competing interest}
The authors declare that they have no known competing financial interests or personal relationships that could have appeared to influence the work reported in this paper.

\section*{Data availability}
The simulation data, the board captures and the analysis scripts that support the findings of this article are not publicly available at this time because the design is part of an ongoing instrument development. They can be obtained from the corresponding author upon reasonable request.

\section*{Declaration of generative AI and AI-assisted technologies in the manuscript preparation process}
During the preparation of this work the authors used Claude (Anthropic) to assist in drafting and revising the text and in typesetting the manuscript, including the preparation of the figure sources from the authors' data. After using this tool, the authors reviewed and edited the content as needed and take full responsibility for the content of the published article.

\end{document}